\documentclass[]{spie}  

\usepackage{amsmath,amsfonts,amssymb}
\usepackage{graphicx}
\usepackage{gensymb}
\usepackage[colorlinks=true, allcolors=blue]{hyperref}

\title{DM/WFS mis-registration tracking: Implementation and on-sky validation of SPRINT at LBT}

\author[a]{Ben Buky}
\author[b,c]{C\'{e}dric Ta\"{i}ssir H\'{e}ritier}
\author[d]{Fabio Rossi}
\author[e]{Juan Carlos Guerra}
\author[a]{Charlotte Z. Bond}
\author[a]{Noah Schwartz}
\author[d]{Guido Agapito}
\author[d]{Enrico Pinna}
\author[e]{Sam Ragland}
\author[b,c]{Jean-Fran\c{c}ois Sauvage}
\affil[a]{UK Astronomy Technology Centre, United Kingdom }
\affil[b]{Aix Marseille Univ, CNRS, CNES, LAM, Marseille, France}
\affil[c]{DOTA, ONERA, 13330, Salon-de-Provence, France}
\affil[d]{ INAF – Osservatorio Astrofisico di Arcetri, Italy}
\affil[e]{Large Binocular Telescope Observatory, USA}

\authorinfo{Further author information: (Send correspondence to B. Buky)\\E-mail: ben.buky@stfc.ac.uk}

\begin{document} 
\maketitle

\begin{abstract}
The advent of telescopes with an integrated deformable mirror (DM) presents new challenges for adaptive optics (AO) systems. The alignment between the DM and wavefront sensor (WFS) is expected to regularly evolve during operations due to their large separation. Without tracking and correction, these mis-registrations between the DM and WFS lead to loop instability, preventing diffraction limited performance from being realised. SPRINT\cite{heritier2021} provides an approach to track these mis-registrations during observations. Rotation, shift, and magnification mis-registrations can all be recovered. The Large Binocular Telescope (LBT) currently lacks an operational solution for tracking these mis-registrations, while SPRINT has been selected as the baseline approach for several instruments on the forthcoming Extremely Large Telescope (ELT). We report on the implementation of SPRINT into the LBT real time computer and present experimental results from both daytime and on-sky testing to validate the method.
\end{abstract}


\section{INTRODUCTION}
\label{sec:intro}  

Most major ground-based telescopes now use adaptive optics (AO) to improve the resolution, contrast, and image quality of their observations. An increasing number of these telescopes are embedding AO within the telescope itself, through the use of an integrated deformable mirror (DM)\cite{ELT_M4,LBT_ASM,AOF}. This leads to a large baseline between the DM and wavefront sensor (WFS), often with moving components in the optical path. As a consequence, the relative positions of the DM actuators as seen by the WFS evolves throughout operations. These mis-registrations threaten AO loop performance and stability, jeopardising the quality of the scientific observations. 

\begin{figure} [t]
   \begin{center}
   \begin{tabular}{c} 
   \includegraphics[height=5.1cm]{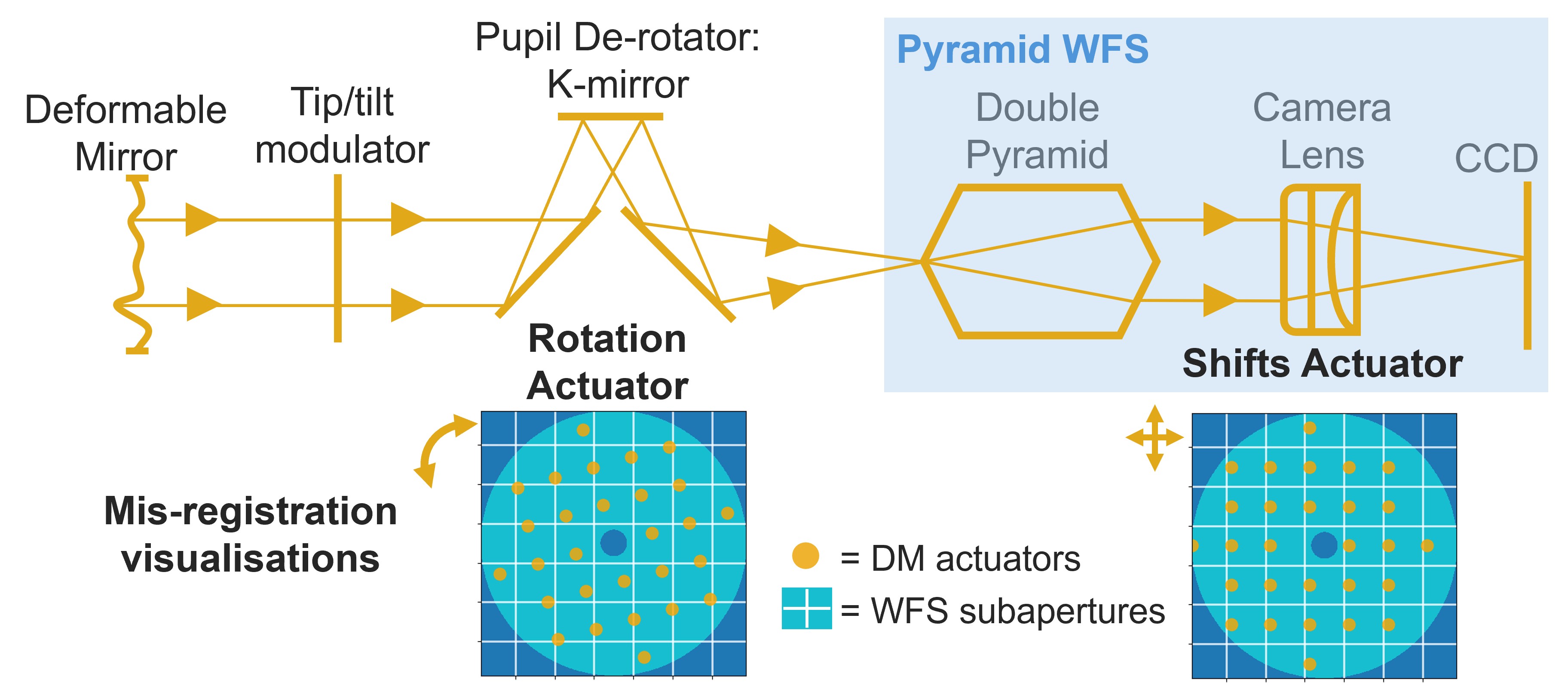}
   \end{tabular}
   \end{center}
   \caption[soul] 
   { \label{fig:soul} 
A simplified optical layout of the SOUL instrument at LBT. The pupil de-rotator and camera lens allow for rotation and shift mis-registrations to be injected and corrected. Visualisations of these types of mis-registrations are provided along the base of the figure.}
   \end{figure}

The primary types of mis-registration\cite{Berdeu_2024}, and the ones studied in this work, are rotation (clocking), and shifts in X and Y. Fig. \ref{fig:soul} contains visualisations of both types. These are the most dynamic mis-registrations during observations. Magnification and anamorphosis effects can also be tracked by the SPRINT method\cite{heritier2021} but have not been implemented here.

High order modes are the most sensitive to mis-registrations\cite{heritier2021}. As AO systems push to correct ever larger numbers of modes, the challenge posed by mis-registrations will only increase. The upcoming Extremely Large Telescope (ELT) has a pre-focal integrated DM\cite{ELT_M4}, M4, and instruments such as HARMONI will be correcting up to 4000 modes\cite{harmoni}. Tracking and correcting mis-registrations during operations will be crucial to achieving diffraction limited performance. 

One existing telescope with an integrated DM is the Large Binocular Telescope (LBT), which has an adaptive secondary mirror (ASM)\cite{LBT_ASM}. This is the pupil stop of the optical system. The SOUL instrument\cite{SOUL} at LBT optically corrects for shift mis-registrations using the WFS camera lens. A control loop measures the centroid positions of the pyramid pupils on the WFS detector and operates the camera lens x-y axes. This works to minimise the difference between the measured positions and those of the calibrated pupils. The K-mirror de-rotator can be used to correct rotation mis-registrations but no method is currently in place to track these during operations. A simplified optical layout of SOUL is found in Fig. \ref{fig:soul}. Implementing a mis-registration tracking approach which can handle both shift and rotation offers performance and stability benefits for LBT. The observatory is also an ideal testbed for ELT instruments as both telescopes have integrated DMs and AO instruments using pyramid wavefront sensors. 

SPRINT\cite{heritier2021} has been introduced previously as a method for tracking mis-registrations during observations. It relies on a pseudo-synthetic model of the AO system to infer the mis-registration parameters from WFS measurements. These measurements contain the response to a small, deliberately injected periodic perturbation on the DM, making SPRINT an inherently invasive technique. The perturbation has a small enough amplitude to ensure a near imperceptible impact on the science. In this work, we present the adapted version of SPRINT implemented at LBT (Sec. \ref{sec:SPRINT@LBT}) and results from recent daytime (Sec. \ref{sec:daytime}) and on-sky (Sec. \ref{sec:on-sky}) testing to validate the method. 

\section{The LBT SPRINT implementation}
\label{sec:SPRINT@LBT}

The original SPRINT algorithm is iterative as both gain variations and mis-registrations are estimated\cite{heritier2021}. Gain variations between the calibrated interaction matrix and observed signal arise from the mis-registrations themselves and the optical gain effects inherent to the pyramid WFS. From one iteration to the next the working interaction matrix of the pseudo-synthetic model of the system is updated using the OOPAO\cite{oopao} software. To reduce the size of the code used to implement SPRINT at LBT, a single iteration version of the algorithm has been developed. This removes the requirement of using the OOPAO package (using native Python instead), but reduces the accuracy of the estimation as it assumes no gain variations are present.

\begin{figure} [t]
   \begin{center}
   \begin{tabular}{c} 
   \includegraphics[height=6cm]{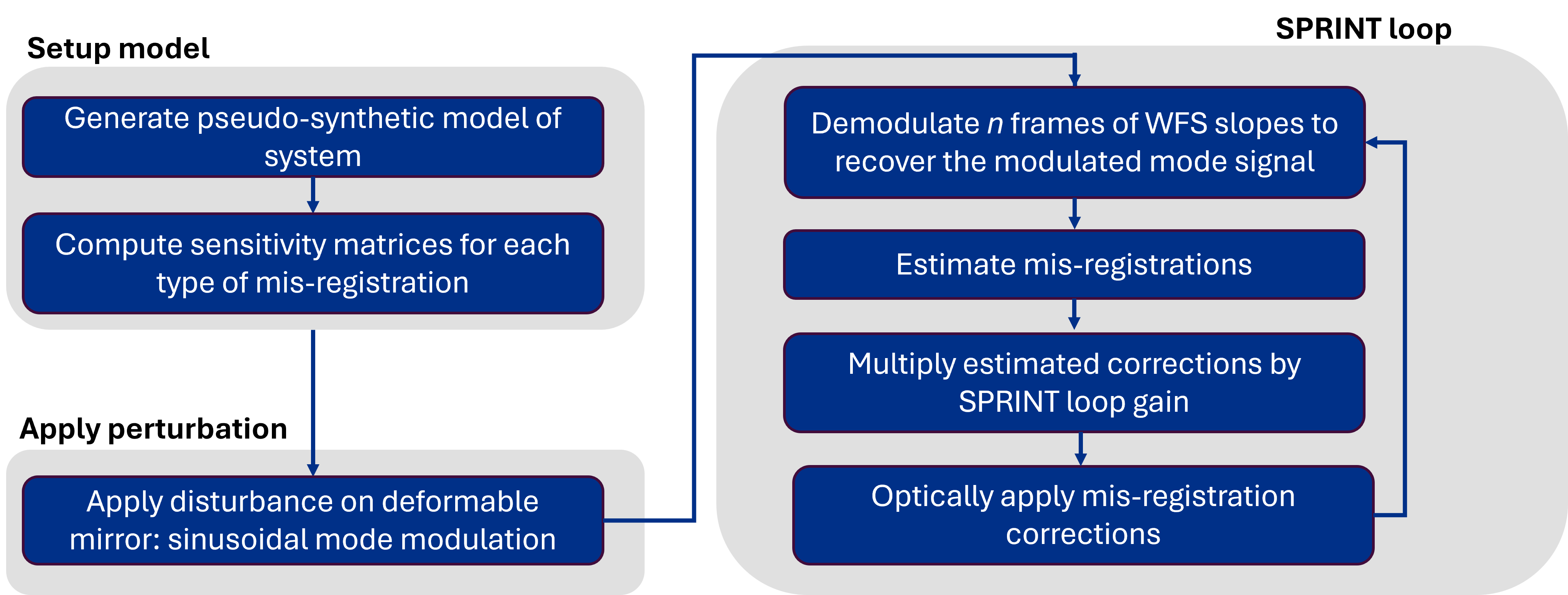}
   \end{tabular}
   \end{center}
   \caption[flowchart] 
   { \label{fig:flowchart} 
The key components of the SPRINT implementation at LBT. The model is set up offline, whilst the perturbation and SPRINT loop are continually running during observations.}
   \end{figure}

Fig. \ref{fig:flowchart} illustrates how this adapted version of SPRINT is implemented at LBT. SPRINT is now an additional auxiliary loop at the telescope. As with all LBT software, the code is adapted for python 2.7. The algorithm requires a pseudo-synthetic model of the AO system. This model is tuned using the current interaction matrix and is used for the generation of mis-registration sensitivity matrices. These encode the response of the system to different forms of mis-registration. The matrices are generated offline before being passed to the telescope. 

The disturbance injected on the LBT ASM is a sinusoidal modulation of Karhunen-Lo\`eve (KL) mode 30 at a temporal frequency of 80 Hz. This perturbation is already applied on the ASM for optical gain tracking\cite{SOUL_gopt, gopt}. No new disturbances are applied for SPRINT and no tuning has been conducted to define the best disturbance for mis-registration tracking. SPRINT is solely using data that is already available.

To generate the input signal for SPRINT, \textit{n} frames of WFS slopes are demodulated. A pixel by pixel demodulation is required to produce the signal of KL mode 30. This is found to be extremely sensitive to the presence of skipped frames in the AO telemetry. This is mitigated against by linearly interpolating the skipped frames and conducting a scan over possible demodulation frequencies around 80 Hz. The frequency with the greatest contrast in the resulting demodulated signal is taken as the optimal result. 

Using the demodulated signal, SPRINT obtains the mis-registration estimates. The target accuracy is to be within 10\% of a subaperture or 0.5\degree\ of the true value, for shift and rotation respectively. These estimates are multiplied by the SPRINT loop gain, the default is 0.6, before being optically applied to the system as mis-registration corrections. The loop repeats throughout an observation to maintain the WFS/DM registration. A graphical user interface has been created for the SPRINT loop. This plots the recent estimates and allows the loop to easily be turned on and off.

\section{Daytime testing results}
\label{sec:daytime}

\begin{table}[b]
\caption{The system parameters used for conducting daytime tests.} 
\label{tab:params}
\begin{center}       
\begin{tabular}{|l|c|} 
\hline
\rule[-1ex]{0pt}{3.5ex}  AO loop frequency & 1700 Hz  \\
\hline
\rule[-1ex]{0pt}{3.5ex}  Seeing of turbulence played on ASM & 1"   \\
\hline
\rule[-1ex]{0pt}{3.5ex}  KL mode perturbed & 30  \\
\hline
\rule[-1ex]{0pt}{3.5ex}  Perturbation amplitude & 10-15 nm   \\
\hline
\rule[-1ex]{0pt}{3.5ex}  Perturbation frequency & 80 Hz   \\
\hline
\rule[-1ex]{0pt}{3.5ex}  SPRINT loop gain & 0.6  \\
\hline
\rule[-1ex]{0pt}{3.5ex}  Frames per SPRINT estimate (\textit{n}) & 500 or 1000  \\
\hline
\end{tabular}
\end{center}
\end{table} 

\begin{figure} [t]
\begin{center}
\begin{tabular}{c} 
\includegraphics[height=5.1cm]{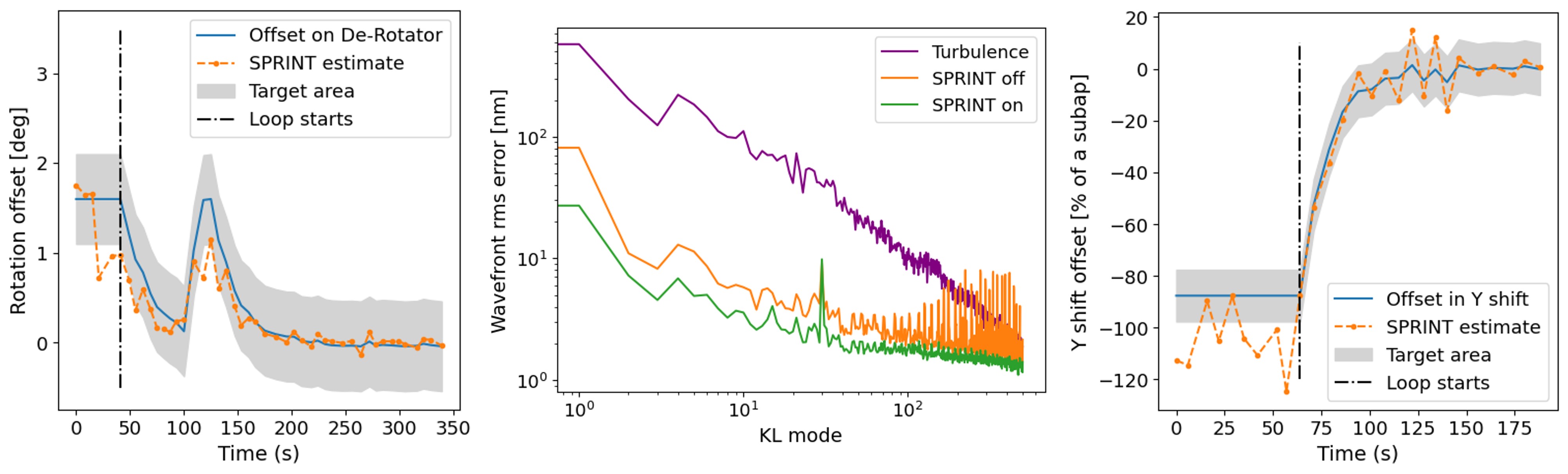}
\end{tabular}
\end{center}
\caption[daytime] 
{ \label{fig:daytime1} 
Results from injecting rotation (left) and y-shift (right) mis-registrations in daytime. The estimates obtained by SPRINT are plotted alongside the true offset values for that mis-registration parameter. Black dashed lines denote when SPRINT begins controlling the offset values. The central plot contains the modal PSD for rotation mis-registration testing.}
\end{figure} 

Table \ref{tab:params} contains the AO system parameters used for testing the SPRINT loop during daytime. Turbulence is played on the ASM to simulate atmospheric turbulence for a seeing of 1". Mis-registrations are injected into the system whilst the SPRINT loop is off (this means SPRINT is measuring the mis-registrations but no corrections are being applied). The loop is then closed to see if SPRINT returns the system to its nominal configuration, and can maintain a steady state there. Fig. \ref{fig:daytime1} contains the results of this test for rotation and y-shift mis-registrations. In both cases, SPRINT is using 500 frames per estimate and the perturbation has an amplitude of 10 nm. 

SPRINT successfully tracks the mis-registrations and returns the system to its nominal configuration for both rotation and y-shift. For the case of rotation, we see that SPRINT initially drives the system towards zero before another mis-registration is applied. This is identified by the algorithm, and the system is once again driven towards its nominal state. For both mis-registration types, the system remains stable once reaching zero. 

The modal PSD plot for the rotation case is found in the central plot of Fig. \ref{fig:daytime1}. In the 'SPRINT off' case, the system is mis-registered and we see how mis-registrations disproportionately affect high order modes. When SPRINT is turned on we see a decrease in wavefront error for all modes, but this is especially noticeable for modes $>$ 100. The spike in wavefront error in the middle of the plot is mode 30. This is the perturbation being applied to the ASM.

\begin{figure} [b]
\begin{center}
\begin{tabular}{c} 
\includegraphics[height=6cm]{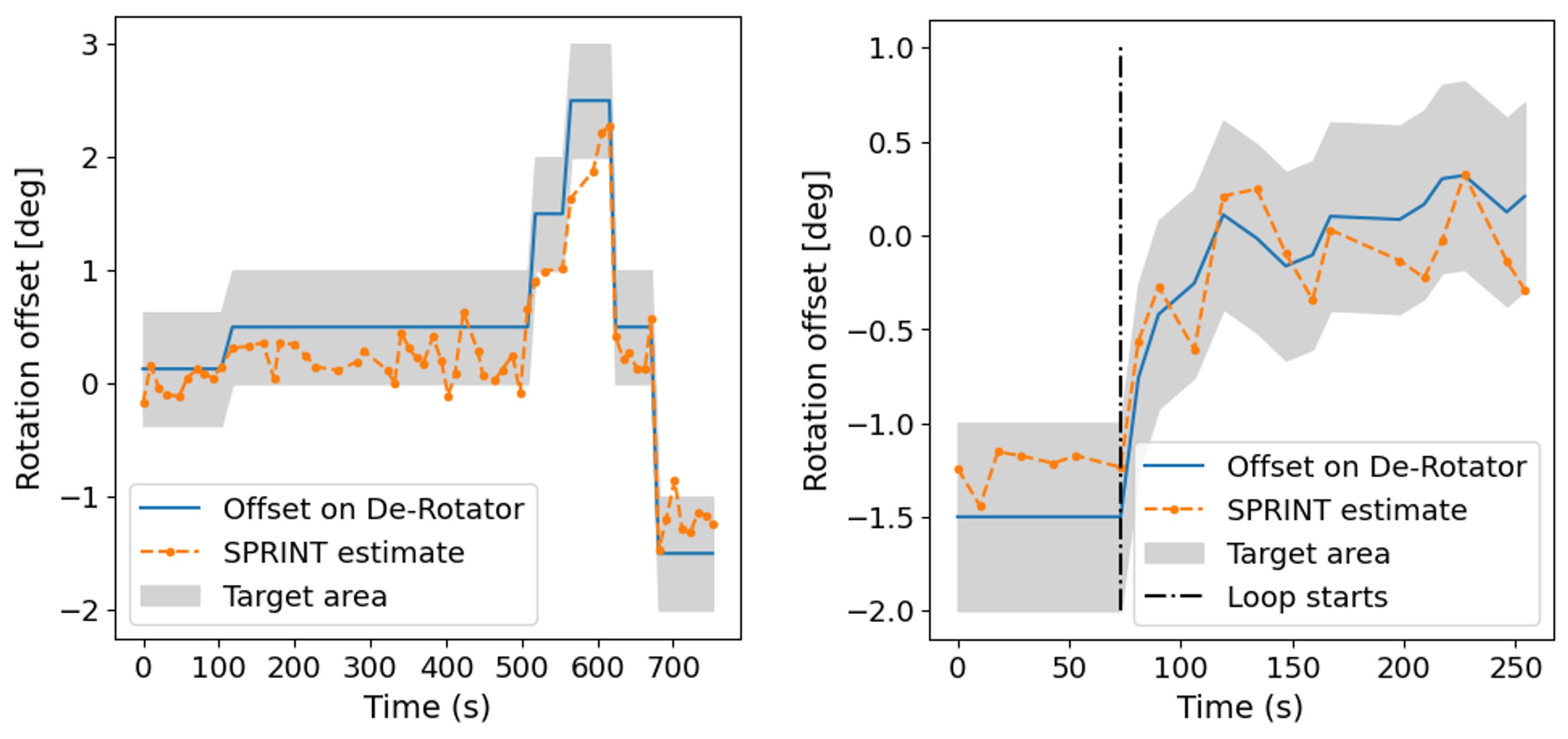}
\end{tabular}
\end{center}
\caption[tt] 
{ \label{fig:tt_only} 
SPRINT rotation tracking results in daytime with the AO loop closed only on tip and tilt. In the left plot SPRINT is measuring the mis-registrations but not applying corrections. In the right plot, SPRINT is used to correct a mis-registration in the system.}
\end{figure}

Mis-registrations are also injected into the system with the AO loop closed only on the tip and tilt modes. Here the perturbation amplitude is increased to 15 nm. Using SPRINT in this way enables mis-registrations to be obtained whilst bootstrapping a system. This enables an AO loop to be closed with its full number of modes even when starting from a severely mis-registered state. This will be crucial for the ELT. The left plot in Fig. \ref{fig:tt_only} is SPRINT tracking rotation mis-registrations in these extreme conditions. In this plot SPRINT is measuring the mis-registrations in open loop. In the right hand plot the SPRINT loop is closed meaning corrections are sent to the de-rotator. We see SPRINT recovers the system back to its nominal state. 

\section{On-sky testing results}
\label{sec:on-sky}

\begin{figure} [t]
\begin{center}
\begin{tabular}{c} 
\includegraphics[height=5.15cm]{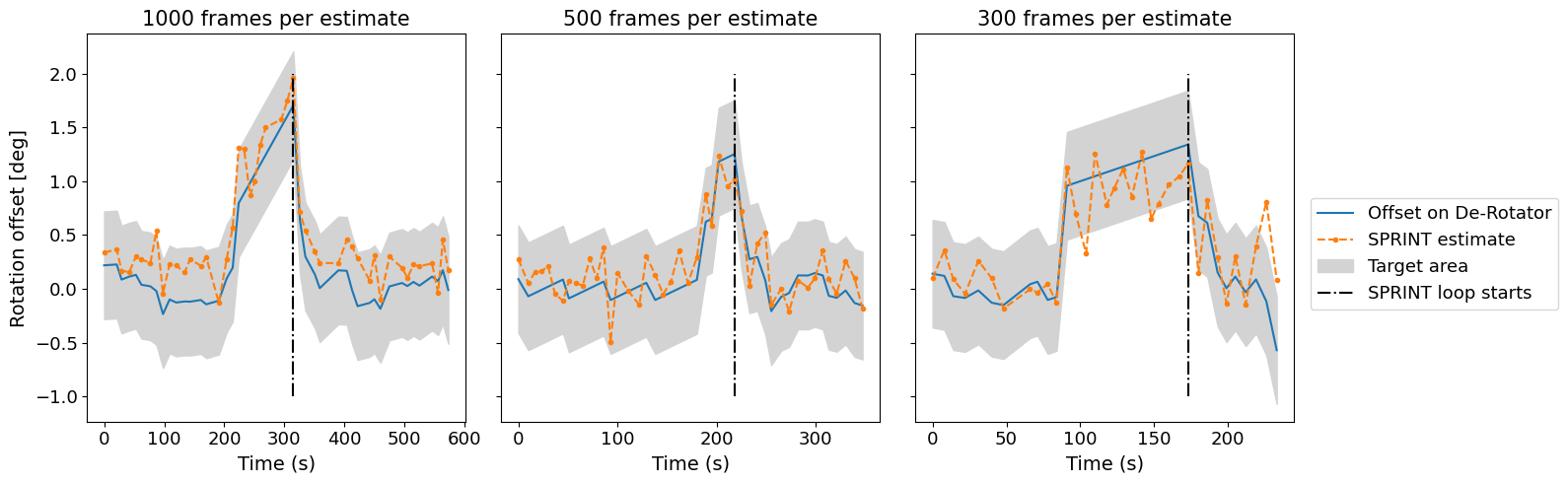}
\end{tabular}
\end{center}
\caption[tt] 
{ \label{fig:onsky_rot} 
On-sky SPRINT results from the injection of rotation mis-registrations. The number of frames per SPRINT estimate is varied across each plot. In each case, the SPRINT estimates, offset on the de-rotator, and time when SPRINT begins controlling the de-rotator are plotted.}
\end{figure} 

\begin{figure} [b]
\begin{center}
\begin{tabular}{c} 
\includegraphics[height=6cm]{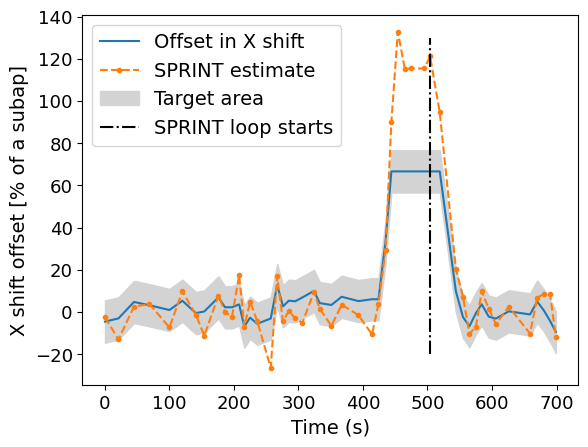}
\end{tabular}
\end{center}
\caption[tt] 
{ \label{fig:onsky_shift} 
On-sky SPRINT results from the injection of a shift mis-registration. The SPRINT estimates are plotted alongside the true x-shift offset on the camera lens.}
\end{figure} 

For on-sky testing, the system parameters are the same as for daytime (see Table \ref{tab:params}) with the following exceptions: the perturbation amplitude is now fixed at 10 nm, and tests are conducted using 300, 500, and 1000 frames per SPRINT estimate. The atmospheric seeing during the tests ranges from 1 - 1.7" and the AO loop is closed on a guide star of magnitude 9. As well as mis-registrations, SPRINT is also correcting for the impact of sky rotation on the guide star orientation. This is not an intended application of SPRINT but we find it absorbs this additional offset into its rotation estimates with no issues.

Similar tests to those conducted in daytime are applied on-sky. This includes the injection of rotation and shift mis-registrations. Rotation mis-registration results for different numbers of frames per estimate are found in Fig. \ref{fig:onsky_rot}. The additional rotation offsets being applied to correct for sky rotation are subtracted from the \textit{Offset on De-Rotator} values for clarity. This means the nominal configuration is 0\degree\ at all times. In all cases we see SPRINT successfully tracks the injected mis-registration, including as the value continues to rise due to sky rotation following the initial injection. No sky rotation corrections are being sent during this time, hence the increasing value. Once SPRINT begins sending commands to the de-rotator the system is returned to its nominal configuration in all cases. More than 95\% of SPRINT rotation estimates in Fig. \ref{fig:onsky_rot} are within the target range. No significant changes in Strehl or modal PSD are observed in these tests. This is most likely due to conservatively small mis-registrations being applied to ensure loop stability is not affected. 

The benefit of using more frames per estimate is clear in Fig. \ref{fig:onsky_rot}. Using 1000 frames leads to the fastest return back to the nominal setup but otherwise 1000 and 500 frames have similar accuracy. However, when we decrease to 300 frames the accuracy worsens and the estimations display larger amounts of variation. 

The 500 frames case in Fig. \ref{fig:onsky_rot} differs in how the system is controlled before the rotation injection. In this case, the standard LBT sky rotation tracking is used, leading to the saw-tooth pattern observed in the \textit{Offset on De-Rotator} values. For the other cases, SPRINT is controlling the system until the mis-registration injection is applied, at which point it is temporarily turned off until the time marked in the plots.

The results from the injection of a shift mis-registration whilst SPRINT used 1000 frames per estimate are seen in Fig. \ref{fig:onsky_shift}. SPRINT recovers the mis-registration but overestimates the initial injection by a factor of two. This behaviour is not observed in the daytime results. Work is ongoing to understand the origin of this factor of two. 

An attempt was made to repeat the test of using SPRINT when the AO loop is closed only on the tip and tilt modes. This coincided with an increase in atmospheric seeing to 1.7", making for challenging conditions. The results are inconclusive and further work is required.  

\begin{figure} [t]
\begin{center}
\begin{tabular}{c} 
\includegraphics[height=5.2cm]{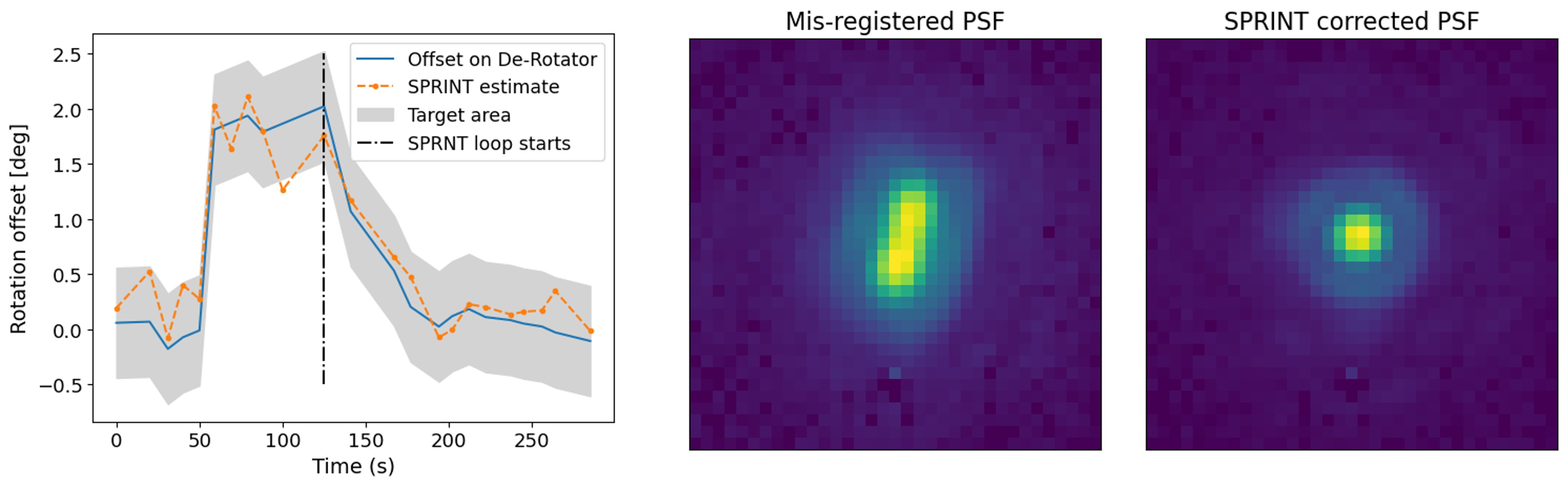}
\end{tabular}
\end{center}
\caption[tt] 
{ \label{fig:onsky_unplanned} 
On-sky SPRINT results from an unplanned rotation mis-registration event. The left plot is the mis-registration estimates and de-rotator values as SPRINT recognises the mis-registration and is turned on to correct for it. The right plots demonstrate the PSF improvement seen by LUCI (near-infrared spectro-imager) due to the correction from SPRINT.}
\end{figure}

During on-sky testing, a large unintended rotation mis-registration was observed. SPRINT was able to detect and correct for this. Some data was gathered when this occurred and this is seen in Fig. \ref{fig:onsky_unplanned}. In the left plot we see SPRINT immediately recognises the mis-registration and is able to return the system to its nominal configuration once it begins controlling the de-rotator. The right plots demonstrate the corresponding PSF improvement as seen by the LUCI near-infrared spectro-imager. An improvement in strehl ratio from 27\% to 67\% is observed. This is an extreme case but it confirms the accuracy of SPRINT outside of planned tests.  

\section{Conclusions and Next Steps}

We have successfully demonstrated the first on-sky validation of SPRINT. Notably, this result was obtained without optimising the injected DM perturbation and using a reduced-complexity implementation of the algorithm with lower nominal accuracy, developed under limited testing opportunities. In both daytime and on-sky tests, SPRINT is able to accurately identify and correct for shift and rotation mis-registrations. The system is returned to its nominal configuration in all cases. This achievement represents an important milestone for SPRINT and provides strong confidence in its readiness for implementation on ELT-class instruments.

In daytime we successfully demonstrate the use of SPRINT when an AO loop is closed only on the tip and tilt modes. This is a significant result in the context of using SPRINT for bootstrapping purposes at the ELT. Further work is required to validate this on-sky. 

We found a factor of two difference between the SPRINT shift estimates and true values on-sky. Additional tests are required to understand if this is a repeatable phenomena and why we do not see this in daytime.

The next steps for SPRINT at LBT are an on-sky validation with the AO loop closed only on the tip and tilt modes and tests to resolve the factor of two discrepancy in shift estimates. We will then seek to expand the implementation of SPRINT to both sides of the telescope and to all binning modes. In the case of ELT instruments, we are ready to work with instrument consortia on the implementation of SPRINT within their systems.

\acknowledgments 
 
The authors would like to thank the LBTO for their time and support in allowing us to test the algorithm. This research is funded by the STFC’s Centre for Innovation (CfI). The LBT is an international collaboration among institutions in the United States and Europe. LBT Corporation Members are: The University of Arizona on behalf of the Arizona Board of Regents; Istituto Nazionale di Astrofisica, Italy; The Ohio State University, representing OSU, University of Notre Dame, University of Minnesota and University of Virginia.

\bibliography{report} 
\bibliographystyle{spiebib} 

\end{document}